\documentclass[12pt,preprint]{aastex}

\begin{document}
\slugcomment{ApJL accepted 2011.03.29}
\title{Heating and Cooling Protostellar Disks}

\author{S. Hirose\altaffilmark{1} and N. J. Turner\altaffilmark{2}}
\altaffiltext{1}{Institute for Research on Earth Evolution, Japan
  Agency for Marine-Earth Science and Technology, 3173-25 Showamachi,
  Kanazawa-ku, Yokohama, Kanagawa 236-0001, Japan}
\altaffiltext{2}{Jet Propulsion Laboratory, California Institute of
  Technology, Pasadena, California 91109, USA;
  neal.turner@jpl.nasa.gov}

\begin{abstract}
  We examine heating and cooling in protostellar disks using 3-D
  radiation-MHD calculations of a patch of the Solar nebula at 1~AU,
  employing the shearing-box and flux-limited radiation diffusion
  approximations.  The disk atmosphere is ionized by stellar X-rays,
  well-coupled to magnetic fields, and sustains a turbulent accretion
  flow driven by magneto-rotational instability, while the interior is
  resistive and magnetically dead.

  The turbulent layers heat by absorbing the light from the central
  star and by dissipating the magnetic fields.  They are
  optically-thin to their own radiation and cool inefficiently.  The
  optically-thick interior in contrast is heated only weakly, by
  re-emission from the atmosphere.  The interior is colder than a
  classical viscous model, and isothermal.

  The magnetic fields support an extended atmosphere that absorbs the
  starlight 1.5~times higher than the hydrostatic viscous model.  The
  disk thickness thus measures not the internal temperature, but the
  magnetic field strength.  Fluctuations in the fields move the
  starlight-absorbing surface up and down.  The height ranges between
  13\% and 24\% of the radius over timescales of several orbits, with
  implications for infrared variability.

  The fields are buoyant, so the accretion heating occurs higher in
  the atmosphere than the stresses.  The heating is localized around
  current sheets, caused by magneto-rotational instability at lower
  elevations and by Parker instability at higher elevations.  Gas in
  the sheets is heated above the stellar irradiation temperature, even
  though accretion is much less than irradiation power when
  volume-averaged.  The hot optically-thin current sheets might be
  detectable through their line emission.
\end{abstract}


\section{INTRODUCTION}

Measurements of the temperatures, densities and flow velocities inside
protostellar disks are vitally important for understanding planet
formation.  New probes of the conditions are made possible by the
detections of infrared line emission from molecules including H$_2$O,
OH, CO, CO$_2$, C$_2$H$_2$ and HCN, in the central 3~AU of the disks
around T~Tauri stars \citep{nc03,ct04,cn08,sp08,ps10}.  However
interpreting the line emission requires some knowledge of the power
source.  The line excitation temperatures range from a few hundred to
a few thousand Kelvin, and can be reached in gas near 1~AU through
accretion heating, if the stresses in the disk atmosphere approach the
local gas pressure \citep{gn04}.

The accretion stresses in T~Tauri disks likely come from
magneto-rotational turbulence \citep{bh98}.  Magnetic fields couple
only to gas that is sufficiently ionized, and while much of the disk
is too cold for thermal ionization, the top and bottom layers are
ionized by the X-rays from the young star \citep{ig99,in06a}.  The
turbulent, magnetically-active surface layers are optically-thin, and
thus produce line emission directly \citep{bg09}.

Another important heat source is the light from the central star.  The
starlight enters the disk at a grazing angle, and so is absorbed high
in the atmosphere, well above the surface of optical depth unity for
the disk's own longer-wavelength continuum emission \citep{cg97}.  The
stellar illumination dominates the atmospheric heating at 1~AU in
traditional viscous models, while the accretion dominates in the
interior, at typical accretion rates around
$10^{-8}$~M$_\odot$~yr$^{-1}$ \citep{dh07}.

Here we report first-principles radiation-MHD calculations of the
heating and cooling, treating the conversion of gravitational to
magnetic energy through magneto-ro\-ta\-tion\-al instability, the
dissipation of the magnetic energy as heat in the gas, and the
emission and escape of the disk radiation.  The calculations build on
recent work by \cite{fk10} in including the stellar irradiation
heating and allowing distinct gas and radiation temperatures.  We also
choose a much lower surface density of $1\,000$~g~cm$^{-3}$, similar
to that at 1~AU in the minimum-mass Solar nebula \citep{h81}, so that
temperatures are too low for collisional ionization.  The ionization
is dominated by the stellar energetic photons, which are absorbed in
the atmosphere, leaving a magnetically-decoupled interior dead zone.

\section{METHODS}

We solve the equations of radiation-MHD using the ZEUS code
\citep{sn92b}, conserving total energy following \cite{hk06}.  The
direct starlight is treated by explicitly solving the transfer
equation, while the disk's own emission is treated using the
flux-limited diffusion module of \cite{ts01}.  The gas and radiation
energy equations are advanced through time in fully-implicit fashion
using a generalized Newton-Raphson method \citep{sm92} with multigrid
solver.  Radiation emission, absorption and diffusion are computed
simultaneously, as needed for accuracy when two of them almost cancel.
We use opacities for homogeneous spherical silicate grains of normal
iron content, from \cite{sh03}.  Opacities for the direct starlight
are averaged over the stellar spectrum, while for the disk's radiation
we use the Planck and Rosseland means at the local temperature.

The star has one-half the mass and twice the radius of our Sun, and
the spectrum of a $4\,000$~K blackbody.  The starlight enters the
domain top and bottom at a grazing angle, 0.05~radians $\approx
3^\circ$ from the horizontal.  The heating is computed each timestep
by solving the transfer equation in the horizontally-averaged density
profile.  We neglect the small amount of absorption that would occur
outside the MHD domain.

The Ohmic resistivity varies in space and time, and is found by
interpolating in a precomputed lookup table whose three axes are the
density, temperature and local ionization rate.  The table holds
equilibrium solutions of the simplified \cite{in06a} recombination
network including reactions on grains (their Model~4).  The grain size
of 0.1~$\mu$m is chosen to give almost the same geometric
cross-section per unit mass as the particle size distribution used for
the opacities.  The two differ by less than 15\%.  The ionization rate
due to the stellar X-rays and long-lived radionuclides is computed
from the horizontally-averaged overlying mass column using the
prescriptions in \cite{ts08}, taking an X-ray luminosity $2\times
10^{30}$~erg~s$^{-1}$ and a long-lived radionuclide ionization rate
$1.4\times 10^{22}$~s$^{-1}$ at nominal dust abundance.  The magnetic
diffusivity is inversely proportional to the resulting electron
fraction \citep{bb94} and is very high in the weakly-ionized disk
interior, reaching $6\times 10^{24}$~cm$^2$~s$^{-1}$ (or if
short-lived $^{26}$Al were present, $2\times 10^{21}$).  To avoid
short timesteps, we cap the diffusivity, limiting it to no more than
$5\times 10^{16}$~cm$^2$~s$^{-1}$.  This is below the threshold of
$5\times 10^{20}$ for diffusion to overcome the shearing of radial
fields under differential rotation \citep[eq.~1 of][]{ts08}, so weak
magnetic activity is expected in the dead zone.  However this is
probably unimportant for the active layers, as we obtain similar MHD
results with a cap ten times higher.

T~Tauri disk atmospheres appear dust-depleted by factors of $10^{1-4}$
compared with the interstellar medium, based on their mid-infrared
colors \citep{fh06,fw09}.  Atmospheric depletion is consistent with
modeling of some protostellar disks over wider wavelength ranges,
incorporating spatially-resolved data \citep{dm03,pp08}.  Furthermore,
the concentration of solid material in the interior is helpful for
planet formation.  We therefore assume a dust-to-gas mass ratio
$10^{-4}$ and reduce all dust effects 100~times below nominal,
including the opacities, the cross-sections for grain surface
reactions, and the radionuclide decay ionization.  The dust is
well-mixed, and thermally coupled to the gas so they share a single
temperature, which is allowed to differ from the local radiation
temperature.  The gas has mean molecular weight 2.3 and an ideal
equation of state with $\gamma=1.4$.  A density floor of $10^{-5}$
times the initial midplane density is applied to prevent high Alfven
speeds from halting the calculations.  We checked the results are
basically unchanged with a density floor ten times lower.  Flow
velocities also are limited to no more than 10~times the shear speed,
$\frac{3}{2}\Omega$ times the domain width.

The domain is a box of size $0.07\times 0.28\times 0.7$~AU along the
radial, orbital and vertical directions, centered in the midplane at
1~AU and divided into $32\times 64\times 320$ grid cells.  The
boundaries are shearing-periodic, periodic, and outflow, respectively.
Like \cite{fk10}, we assume the magnetic fields are vertical on the
top and bottom boundaries.  Results are similar using the standard
ZEUS outflow boundary condition, where the fields on the boundaries
are computed from the extrapolated EMFs.  We also added a small
artificial resistivity near the boundaries for numerical stability
\citep{hk09}.  The disk radiation on the vertical boundaries is placed
in the streaming limit, with flux equal to the product of the
radiation energy density and the speed of light.

Initially the gas is isothermal at 280~K and in hydrostatic balance,
while the magnetic field is the sum of an 0.02-Gauss uniform vertical
component with pressure $3\times 10^5$ times less than the midplane
gas pressure, and a vertical component varying sinusoidally in the
radial direction with maximum pressure $1/270$th the midplane gas
pressure.  Results were similar in a further calculation with the
pressure in the sinusoidal component 100~times less.  We also compared
an ideal-MHD version of our calculation against the results of
\cite{fk10}, obtaining comparable profiles of the plasma beta
parameter, turbulent Mach number and density fluctuations, despite our
lower surface density.

Below we relate the MHD results to the classical viscous, irradiated
disk model with the same surface density and accretion rate.  We
follow \cite{dc98} in solving the equations of vertical hydrostatic
equilibrium, radiative equilibrium in the Eddington approximation, and
thermal balance, together with a transfer equation for the
illuminating starlight.  Scattered radiation is neglected in both the
viscous and MHD calculations.

\section{RESULTS}

\begin{figure}[tb!]
  \epsscale{0.5}
  \plotone{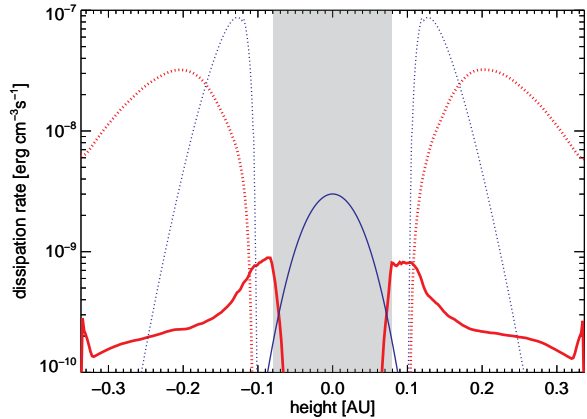}

  \figcaption{Heating profiles in the MHD calculation (red) and the
    classical viscous model with the same surface density and
    accretion rate (blue).  Both have contributions from stellar
    irradiation (dotted) and turbulent or viscous dissipation (solid).
    The turbulent dissipation involves both magnetic and kinetic
    energy.  All are averaged horizontally and over time from 50 to
    200~orbits.  A grey band marks the dead
    zone.  \label{fig:heating}}
\end{figure}

\begin{figure}[tb!]
  \epsscale{0.5}
  \plotone{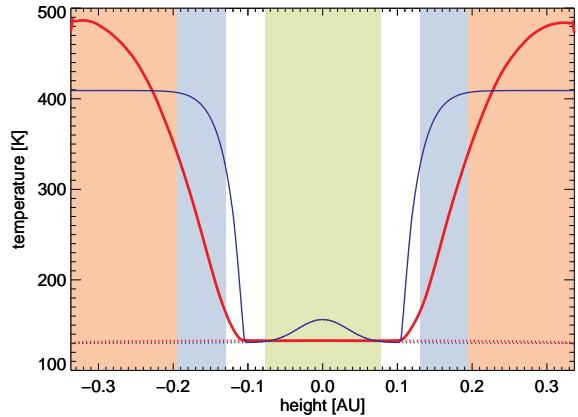}

  \figcaption{Temperature profiles in the MHD calculation (red) and
    corresponding viscous model (blue).  Shadings with the same colors
    indicate the penetration of the starlight to unit optical depth,
    while green shading marks the part of the disk that is opaque to
    its own radiation.  The gas and dust (solid) is hotter than the
    disk radiation (dotted) at low optical depths.  The profiles are
    averaged horizontally, and in time from 50 to 200~orbits.
    \label{fig:temperature}}
\end{figure}

The turbulent layers on the top and bottom are optically-thin to their
own radiation, while the magnetically-dead interior is
optically-thick.  The base of the turbulent layer, defined by unit
Elsasser number $v_{Az}^2/\eta\Omega$ \citep{si01,ts07}, has height
0.08~AU and Rosseland mean optical depth~0.39, so the disk photosphere
falls within the dead zone.  The starlight on the other hand is
absorbed in the turbulent layers, dominating the mean heating above
0.11~AU (Figure~\ref{fig:heating}).  The turbulent layers are more
opaque in the optical than the infrared, and absorb starlight better
than they re-radiate, becoming several times hotter than the disk
radiation passing through them.  The horizontal averaging in the
irradiation prescription yields mean temperatures higher than the true
irradiation temperature of 410~K found in the viscous model.  Below
the disk photosphere the gas and radiation temperatures match.  The
interior, in contrast, receives no direct starlight and is heated by
the re-radiation from the hot surface layers.  The cooling is almost
wholly radiative throughout.  Turbulent advection is negligible, and
unlike the viscous models in \cite{dc98} figure~10, carries no
significant heat to the interior.  The heating and cooling together
yield the temperature profile in Figure~\ref{fig:temperature}.

\begin{figure}[tb!]
  \epsscale{0.5}
  \plotone{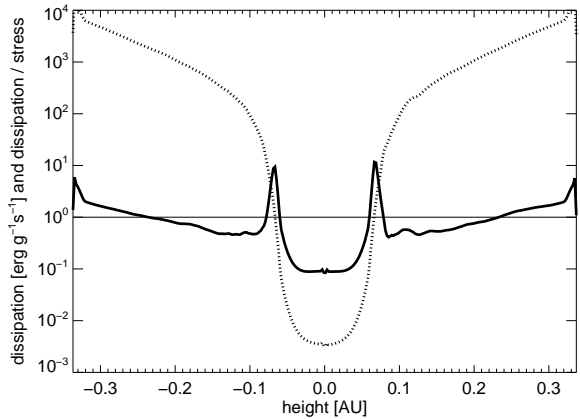}

  \figcaption{Dissipation per unit mass (dotted) and ratio of
    dissipation to accretion stress (solid), versus height.  The
    stress is scaled by the shear rate $\frac{3}{2}\Omega$ so the
    ratio is dimensionless.  Both quantities are averaged horizontally
    and over time from 50 to 200~orbits.
    \label{fig:ratio}}
\end{figure}

The turbulent layers are supported by magnetic forces above 0.12~AU,
while the interior is gas-pressure-supported.  The magnetic
dissipation is the second-largest contribution to the
volume-integrated heating.  The magnetic fields are buoyant and rise
while dissipating, so the accretion heating occurs higher on average
than the accretion stress (Figure~\ref{fig:ratio}).  By contrast, in
the classical model the stress and heating occur at the same place and
time, and both result from an effective viscosity, nature unspecified.
The dissipation rate in our calculation is comparable to the local
pressure $p(z)$ times the shear rate $\frac{3}{2}\Omega$, the ratio
increasing from 0.4 near the turbulent layer base to 2 at the
boundary.  Ratio unity is sufficient to produce CO molecular line
emission in the models of \cite{gn04} at lower columns around
$10^{21}$~cm$^{-2}$.  The dissipation is also high relative to the
stress in two narrow layers near $\pm 0.07$~AU where turbulent
magnetic fields are destroyed on entering the dead zone.  Overall, the
stresses correspond to a mean mass flow rate $1.3\times
10^{-8}$~M$_\odot$~yr$^{-1}$, within the range among classical T~Tauri
stars \citep{mh03}.

\begin{figure}[tb!]
  \epsscale{0.5}
  \plotone{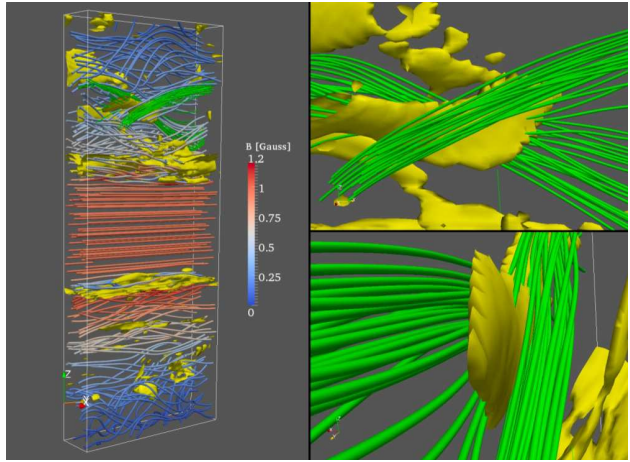}

  \figcaption{Left: Typical snapshot of the magnetic field lines, made
    at 155~orbits.  The field strength in Gauss is indicated by
    colors: stronger fields are red, weaker fields blue.  Inside the
    yellow isosurfaces the magnetic dissipation rate exceeds
    $10^{-9}$~erg~cm$^{-3}$~s$^{-1}$.  Field lines drawn in green are
    those passing through a zone of strong magnetic dissipation at top
    center.  Right: close-up of the green field lines, viewed from the
    star (top) and along the orbit (bottom).  The dissipation is
    concentrated in a current sheet of narrow radial extent,
    separating two counter-undulating bundles of field lines.
    \label{fig:parker}}
\end{figure}

The magnetic field structure varies with height.  Near the base of the
turbulent layers, current sheets are typically close to horizontal,
while high in the atmosphere, current sheets are most often vertical.
Near the base, the current sheets separate magnetic fields moving
radially inward and outward under the magneto-rotational instability,
while higher in the atmosphere, the current sheets separate magnetic
fields lying along adjacent orbits and having out-of-phase vertical
undulations.  The upward slope of one field line annihilates against
the downward slope of the next (Figure~\ref{fig:parker}).  The
undulations come from the Parker instability, whose fastest-growing
modes have large radial wavenumber under Keplerian differential
rotation.  By solving in each grid cell the dispersion relation of
\cite{tp96} eq.~33, describing localized non-axisymmetric disturbances
on toroidal magnetic fields, we have verified that MRI modes ($k_z\gg
k_r$) grow fastest near the base, where the field is weak, while
Parker modes ($k_r\gg k_z$) grow fastest high in the atmosphere, where
the plasma beta is low and magnetic buoyancy strong.  A similar
pattern appears in ideal-MHD models of strongly-magnetized disks
\citep{jl08}.  The Parker modes grow faster than the orbital frequency
in locations where the fields decline steeply with height, including
just below reversals in the sign of the dominant toroidal field, and
along the tops of magnetic flux tubes.  Among the solutions of the
dispersion relation are still faster-growing Parker modes with
azimuthal wavelengths 2-3~times our domain size.  Future experiments
with larger boxes are indicated.  Since the dissipation is
concentrated in thin current sheets (figure~\ref{fig:parker}), it
dominates the heating at some places and times.  Dissipation is
stronger than irradiation in a volume fraction $3.2\times 10^{-4}$,
time-averaging from 50 to 200~orbits and considering the layers above
the disk photosphere ($|z|>0.18$~AU).  Temperatures in the current
sheets are up to 50\% greater than in the gas heated only by
starlight.  Finally, mass leaves the top and bottom boundaries in an
outflow comparable to those in isothermal calculations by \cite{sm10}.

Most of the energy input between 50 and 200~orbits (97\%) comes from
stellar irradiation.  Of the 3\% coming from work done by magnetic
stresses, 52\% is converted directly to heat via magnetic dissipation,
while 26\% is first converted to kinetic energy and then dissipated
via shocks and viscosity.  The remainder crosses the boundaries as
magnetic and turbulent kinetic energy and the gravitational potential
energy of the small amount of escaping gas.  Overall, energy
ultimately leaves the box almost entirely by radiation diffusion.

\begin{figure}[tb!]
  \epsscale{0.75}
  \plotone{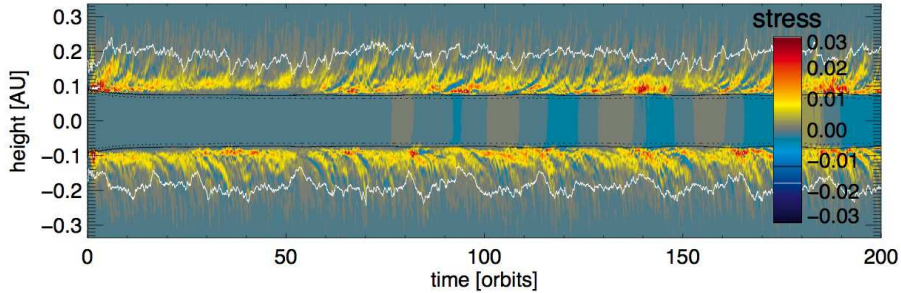}

  \figcaption{Accretion stress vs. height and time, together with the
    surfaces where the starlight is absorbed (white lines) and the
    Rosseland and Planck mean disk photospheres (black dotted and
    solid lines, respectively).  The stress is horizontally-averaged
    and is plotted in units of dyn~cm$^{-2}$.
    \label{fig:tzstress}}
\end{figure}

The surface where the starlight is absorbed moves up and down
irregularly over intervals of several orbits
(Figure~\ref{fig:tzstress}) as fluctuations in the magnetic fields
lead to changes in the density profile.  The fluctuations are
associated with episodic reversals in the toroidal field, which lift
away from the midplane at a rate that increases with height, producing
the butterfly-wing pattern visible in figure~\ref{fig:tzstress}.  A
majority of the volume-integrated accretion heating occurs near these
reversals.  As in previous stratified MRI turbulence calculations
\citep[e.g.][]{ms00,hk06,bh07,fk10}, the density in the
magnetically-dominated atmosphere fluctuates horizontally by more than
a decade, and the velocity dispersion exceeds the sound speed,
approaching the Alfven speed.

\section{DISCUSSION}

Diagnosing conditions in planet-forming disks requires an
understanding of the power sources for the observed line and continuum
emission.  As a step in this direction, we carried out
energy-conserving radiation-MHD calculations of heating and cooling in
a local shearing box.  The starlight that dominates the overall
heating is absorbed in an extended, magnetically-supported atmosphere.
The surface of optical depth unity lies on average 1.5~times higher
than in the corresponding hydrostatic model.  The fraction of the
stellar luminosity intercepted will be larger in proportion, with
consequences for interpreting the spectral energy distributions of
protostellar disks.  The puffy, magnetically-supported atmospheres may
account for the puzzling fact that many young stars have near-infrared
excesses too large to explain with hydrostatic models, as reviewed by
\cite{dm10}.  Greater dust mass fractions than our assumed value of
$10^{-4}$ would raise the opacities, pushing the absorbing surface yet
higher, while also increasing the recombination cross-section, making
the weakly-ionized dead zone thicker.

The accretion flow occurs in the disk atmosphere, not the interior as
assumed in classical viscous models, and the dissipation occurs still
higher (figure~\ref{fig:heating}).  The viscous model's midplane
temperature bump vanishes in the MHD case: relocating the dissipation
to the atmosphere leaves the interior cooler and isothermal
(figure~\ref{fig:temperature}).  The magnetic fields are generated in
an optically-thin turbulent layer and rise buoyantly as they
dissipate.  While the total mass column is $1\,000$~g~cm$^{-2}$, over
half (56\%) of the accretion heat is deposited at columns within
1~g~cm$^{-2}$ of the boundaries, and more than one-fifth (23\%) within
0.1~g~cm$^{-2}$.  The turbulent layer is optically-thin to its own
continuum emission independent of the dust depletion, so long as the
grains control both the opacity and the recombination \citep{bg09}.
Because the accretion power heats optically-thin gas, the resulting
spectral lines can potentially be used to measure the mass flow rate.
The dissipation is spatially inhomogeneous, heating gas locally above
the irradiation temperature despite the starlight dominating the
volume-integrated power.  Even a small disk patch thus produces
emission at a range of temperatures.

Since the disk atmosphere is magnetically-supported, its structure
varies in response to changes in the strength of the fields.  In
particular, the height where the starlight is absorbed moves up and
down by almost a factor of two over timescales of several orbits, with
an amplitude around 10\% of the distance to the star.  The resulting
changes in the stellar illumination of the disk surface could
contribute to the widespread low-amplitude infrared variability among
young stars with disks \citep{ms09,la10}.

To make the calculations feasible, we simplified the situation in
several respects.  Among these, we capped the magnetic diffusivity
well above the threshold for shutting off the MRI, but well below the
value computed from the midplane ionization fraction.  The weak
midplane magnetic oscillations visible in figure~\ref{fig:tzstress}
are an artifact of the reduced midplane diffusivity, and would
disappear if the cap were lifted above the threshold where diffusion
cancels the winding-up of radial fields.  The dead zone stresses are
unimportant however, as they contribute only 2.6\% to the total mean
mass accretion rate.  Also, we treated the stellar irradiation in a
horizontally-averaged fashion and kept the starlight incidence angle
fixed despite the changes in the disk thickness, since the
shearing-box approximation gives no information on the tilt of the
disk surface.  Future work could explore the effects of relaxing these
simplifications.  In addition, our current sheets are artificially
broadened to the grid scale by the numerical resistivity.  The sheets
should be narrower, with still greater accretion heating per unit
volume, and thus greater chances of exceeding the stellar irradiation.
The magnetic dissipation furthermore heats the gas, which may become
hotter than the dust where densities are low enough so heat transfer
by gas-dust collisions is slow.  Since we assumed good gas-dust
thermal coupling, our peak temperatures are lower limits.  Future
modeling should follow \cite{gn04}, \cite{kd04} and \cite{nm05} in
allowing distinct dust and gas temperatures.

\acknowledgements

We were supported by a Grant-in-Aid for Scientific Research (No.\
20340040) from MEXT (S.~H.), by NASA Origins grant 07-SSO07-0044 and
by the Alexander von Humboldt Foundation (N.~J.~T.).  Calculations
were made on the Cray XT4 at the CfCA, National Astronomical
Observatory of Japan, and Altix3700 BX2 at the Yukawa Institute for
Theoretical Physics, Kyoto University.  Part of the research was
carried out at the Jet Propulsion Laboratory, California Institute of
Technology, under a contract with the National Aeronautics and Space
Administration.  We are grateful to Dmitri Semenov for the opacities.
S.~H.\ thanks Kengo Tomida and Takayoshi Sano, and N.~J.~T.\ thanks
John Carr, Hubert Klahr and Joan Najita for discussions.  The project
began at the CPS, Kobe University, under the MEXT Global COE Program,
``Foundation of International Center for Planetary Science''.

\end{document}